# Oxidation State of Cobalt in the $Na_xCoO_{2-\delta} \cdot y\,H_2O$ Superconductor

Maarit Karppinen*, Isao Asako, Teruki Motohashi, Hisao Yamauchi

*Materials and Structures Laboratory, Tokyo Institute of Technology,*

*Yokohama 226-8503, Japan*

*To whom correspondence should be addressed. Fax: (+81)45-924-5365. E-mail: karppinen@msl.titech.ac.jp.

Here we present results of accurate wet-chemical redox analyses, revealing that the oxidation number of cobalt in the newly discovered $CoO_2$-layer superconductor, $Na_xCoO_2 \cdot y\,H_2O$ ($x \approx 0.36$, $y \approx 1.3$) is ~3.46. This value is significantly lower than the one (~3.64) expected on the basis of the value determined for the Na to Co stoichiometry ratio, $x$. The "lower-than-expected" value for the Co oxidation state may be alternatively explained by presence of oxygen vacancies or excess protons. Moreover, the result implies that rather than an electron-doped $Co^{IV}$ lattice the phase should be considered as a hole-doped $Co^{III}$ lattice.

Prof. Maarit Karppinen
Materials and Structures Laboratory
Tokyo Institute of Technology
4259 Nagatsuta, Midori-ku
Yokohama 226-8503, JAPAN
fax:     +81-45-924-5365
tel:     +81-45-924-5333
e-mail:  karppinen@msl.titech.ac.jp

## Introduction

For the last one and half decades unprecedented research efforts have been focused on layered strongly-correlated-electron oxide materials. These efforts were initially inspired by the discovery of high-$T_c$ superconductivity[1] and more recently by discoveries of other spectacular phenomena such as large magnetoresistance[2] and high-efficiency thermoelectricity[3]. Very recently, superconductivity with the superconductivity transition temperature, $T_c$, at 4 ~ 5 K was discovered in $CoO_2$ layers[4] stacked together with layers of $Na^+$ ions and "iced"[5] water molecules. The superconducting phase, $Na_xCoO_2 \cdot yH_2O$ ($x \approx 0.35$, $y \approx 1.3$), is obtained from its sodium-rich parent, $Na_xCoO_2$ ($x \approx 0.7$)[6] through oxidative deintercalation[7] of sodium and subsequent intercalation of water in between each pair of adjacent $CoO_2$ layers. Neither the relatively low value of $T_c$ nor the insufficient chemical stability of the new superconductor predicts good at application market. Nevertheless, the discovery has already gained considerable attention.[8] Here the remarkable point is that the discovery made it possible for us to mirror the yet puzzling high-$T_c$ copper-oxide superconductor against another layered $3d$ transition metal oxide superconductor.

The similarities of the $CoO_2$- and the $CuO_2$-layer superconductors are evident: both are nonstoichiometric and possess crystal structures in which the nonsuperconducting or insulating layers that alternate with the $CoO_2/CuO_2$ layers play a dual role of a "spacing provider" and a "redox controller". On the other hand, an obvious difference is seen in the way the Co-O/Cu-O polyhedra are linked each other: in $Na_xCoO_2 \cdot yH_2O$ (despite the values of $x$ and $y$) the "$CoO_2$ layer" is rather a three-layer block of triangular O, Co and O layers (alternatively the same may be imagined as a thick layer of edge-sharing $CoO_6$ octahedra), whereas copper in the superconducting "$CuO_2$ layer" possesses square-planar coordination with nearly 180° O-Cu-O bond angle. For Cu-based high-$T_c$ superconductors the crucial importance of the oxidation state of Cu in the $CuO_2$ layers was established in an early stage of research, in both switching superconductivity on and off and controlling the value of $T_c$ in the superconductivity regime.[9] For $Na_xCoO_2 \cdot yH_2O$, no direct experimental data have been reported that would enable us to establish the oxidation state of Co in the $CoO_2$ layer. Given the Na content at $x = 0.30 \sim 0.35$, the phase has simply been considered as an electron-doped system of low-spin $Co^{IV}$ with an electron density of 0.30 ~ 0.35 *per* Co atom, *i.e.* an oxidation number of 3.65 ~ 3.70 for Co.[4,10] The main conclusion of the present work - reached by means of accurate wet-chemical redox analysis - is that cobalt in the superconducting $CoO_2$ layer possesses an oxidation number markedly lower than what has been expected, that is, a value even somewhat lower than 3.5. An apparent implication of this is that rather than an electron-doped Co$^{IV}$ lattice the phase should be considered as a hole-doped Co$^{III}$ lattice. For comparison, we also report valence analysis results for a series of nonsuperconducting Na$_x$CoO$_2 \cdot y$H$_2$O samples with varying $x$ and $y$.

## Samples Synthesis and Basic Characterization

A single-phase sample of the γ-Na$_x$CoO$_2$ phase[6] with a nominal cation stoichiometry of $x$ = 0.70 (thereafter referred to as Sample A) was first synthesized as a precursor utilizing a "rapid-heat-up" technique[11] where a powder mixture of Na$_2$CO$_3$ and Co$_3$O$_4$ in an Al$_2$O$_3$ crucible is directly placed in a furnace preheated at 800 °C and fired for 12 hours in air. Deintercalation of Na was performed following the precept given in a previous work,[12] *i.e.* stirring a 500-mg portion of Sample A in 20 ml solution of 6 M Br$_2$ in CH$_3$CN at room temperature for one day. The resultant product was washed with CH$_3$CN several times and then dried in a fume box to obtain a Na-poor, nonhydrated Sample B. Finally, the fully-hydrated, superconducting Na$_x$CoO$_{2-\delta} \cdot y$H$_2$O phase (Sample C) was obtained from Sample B by keeping it in saturated humidity for a few days. To check the reproducibility of the results, "Sample C" was synthesized in triplicate; the three parallel samples turned out to be identical in respect to every piece of characteristics measured.

The precise Na to Co cation ratio was determined for the three samples, A, B and C, by inductively-coupled-plasma atomic-emission spectroscopy (ICP-AES; Hitachi: P-5200) at $x$ = 0.72(2), 0.36(2) and 0.36(2), respectively (Table 1). From the x-ray diffraction patterns for the samples, the $c$-axis lattice parameter was respectively determined at 10.93(1), 11.22(1) and 19.63(1) Å. Bulk superconductivity with $T_c$ = 4.4 K was confirmed for Sample C with a large volume fraction as judged from a zero-field-cooled magnetization curve measured upon heating (starting from 2 K) under an applied magnetic field of 10 Oe in a SQUID magnetometer (Quantum Design: MPMS-XL), see Figure 1. The values presently revealed for $T_c$ (= 4.4 K), Na content ($x$ = 0.36), and $c$-axis parameter (= 19.63 Å) of Sample C well agree with those assigned in the original report by Takada *et al*.[4] for the novel Na$_x$CoO$_2 \cdot y$H$_2$O superconductor, *i.e.* 4 ~ 5 K, 0.35 and 19.62 Å, respectively. The $c$-axis parameter of 10.93 Å for Sample A (with $x$ = 0.72) may be compared to that of 10.88 Å reported for Na$_{0.71}$CoO$_2$.[13]

In order to determine the precise water content of the superconducting phase, we subjected a portion of Sample C to thermogravimetric (TG) annealing in which water accommodated in it is gradually but completely removed. The TG annealing was carried out in a high-sensitivity thermobalance (Perkin Elmer: Pyris 1) by heating a specimen of *ca.* 20 mg in $O_2$ up to 300 $^o$C with a rate of 0.25 $^o$C/min. Owing to the slow heating rate the resultant TG curve should represent a nearly equilibrium state at each temperature. To warrant the result, the experiment was repeated several times. In Figure 2, three representative TG curves are shown to demonstrate the fact that the dehydration process was highly reproducible. Another confirmation of the data is that the present TG curves are almost identical to that reported previously by Foo *et al.*[12]. From Figure 2, the dehydration occurs in several distinct steps by 230 $^o$C resulting in an overall loss of 1.3(1) $H_2O$ molecules *per* formula unit. In other words, we determined the magnitude of *y* at 1.3(1) for Sample C.

We further utilized the TG data given in Figure 2 to consider the dehydration and decomposition characteristics of the $Na_xCoO_2 \cdot yH_2O$ system in more detail. About 130 $^o$C a clear change is seen in the slope of the TG curve indicating that a relatively stable intermediate phase is formed at temperatures somewhat below 130 $^o$C. To obtain a specimen of this phase (Sample D) we annealed a portion of Sample C in air at 120 $^o$C for 1 hour: a single-phase sample was obtained with the *c*-axis lattice parameter at 13.83(1) Å (and the value of *y* at 0.5(1) on the basis of the TG data). Upon further heating beyond 130 $^o$C, the water remained in the original Sample C is lost (in two or three fairly distinguishable steps) by 230 $^o$C. Guided by this observation, 220 $^o$C was selected for the annealing temperature to synthesize (in air for 1 hour) a specimen of the completely dehydrated $Na_{0.36(2)}CoO_{2-\delta}$ phase (Sample E) starting again from another portion of Sample C. For Sample E the *c*-axis parameter was determined at 11.17(1) Å. The further weight loss seen in TG curves above 230 $^o$C (Figure 1) is apparently due to partial reductive decomposition of the completely dehydrated phase of $Na_{0.36(2)}CoO_{2-\delta}$. Accordingly, peaks due to $Co_3O_4$ (besides those for the main phase of $Na_xCoO_{2-\delta}$) were clearly seen in the x-ray diffraction pattern recorded for a sample obtained by annealing a portion of Sample C in air at 280 $^o$C for 1 hour (Sample F). For the $Na_xCoO_{2-\delta}$ phase contained in Sample F the *c*-axis parameter was determined at 11.06(1) Å. In Figure 3 shown are x-ray diffraction patterns for all the six samples, A, B, C, D, E and F.

## Chemical Analysis for the Oxidation State of Cobalt

The oxidation state of cobalt, $V$(Co), was determined for the five single-phase samples, A, B, C, D and E, utilizing two independent wet-chemical redox analysis methods, *i.e.* cerimetric and iodometric titration, which are known to accurately detect the amount of $Co^{III}$ and/or $Co^{IV}$ in layered cobalt oxides.[14] Additionally, a "$Na_{0.75}CoO_{2-\delta}$" sample of the $\gamma$-$Na_xCoO_2$ phase was synthesized (using the same synthesis conditions as for Sample A) for a reference. Note that ICP-AES analysis revealed a cation stoichiometry of $Na_{0.77(2)}CoO_{2-\delta}$ for this sample (Table 1). The titration experiments were carried out by dissolving ~20 mg of the sample in oxygen-freed 3 M HCl solution containing an excess of the reductant, $Fe^{2+}$ (cerimetric titration) or $I^-$ (iodometric titration). After the complete reduction of high-valent Co species to divalent state, the amount of left-over $Fe^{2+}$/liberated $I_2$ was determined volumetrically using $CeSO_4$/$Na_2S_2O_3$ solution as a titrant. The experimental details were as those given elsewhere.[14] For both the methods parallel experiments revealed the value of $V$(Co) with a reproducibility better than ±0.01. Moreover the two independent methods gave highly agreeable results. The values for $V$(Co) as summarized in Table 1 are all average values of more than four parallel experiments. Knowing the precise value of $V$(Co) from the titration analysis as well as the precise Na content, $x$, from ICP-AES analysis and assuming charge neutrality the degree of oxygen nonstoichiometry, $\delta$, was calculated for each sample (Table 1).

The two sodium-rich samples, $Na_{0.77(2)}CoO_{2-\delta}$ (reference) and $Na_{0.72(2)}CoO_{2-\delta}$ (precursor; Sample A), were found nearly stoichiometric in terms of oxygen, *i.e.* $\delta$ was determined at -0.02(1) for the former and at 0.01(1) for the latter. Upon oxidative deintercalation of $Na^+$ ions, Co species get oxidized as expected, *i.e.* $V$(Co) increases from 3.26(1) (for Sample A) to 3.48(1) (for Sample B). On the other hand, intercalation of neutral $H_2O$ molecules into the Na-poor $Na_{0.36(2)}CoO_{2-\delta}$ phase maintains the oxidation state of cobalt essentially constant: cerimetric titrations reveal $V$(Co) at 3.48(1) and iodometric titrations at 3.44(1) for the superconducting $Na_{0.36(2)}CoO_{1.91(1)} \cdot 1.3(1) H_2O$ phase (Sample C). Partial dehydration of Sample C to obtain the $Na_{0.36(2)}CoO_{2-\delta} \cdot 0.5(1) H_2O$ phase (Sample D) does not change the value of $V$(Co) either. For the completely dehydrated $Na_{0.36(2)}CoO_{2-\delta}$ (Sample E), the $V$(Co) value was slightly lower, *i.e.* 3.37(1). This may originate from small amounts (nondetectable by x-ray diffraction) of $Co_3O_4$ impurity with Co at an oxidation state of 2.67. (Sample E was obtained under conditions that are close to those causing decomposition of the Na-deficient $Na_xCoO_{2-\delta}$ phase, *cf.* Sample F.)

**Discussion**

The present study showed that the value of $V$(Co) is essentially the same for the three Na-poor samples, the superconducting $Na_{0.36(2)}CoO_{1.91(1)} \cdot 1.3(1)\,H_2O$ sample (Sample C) and the nonsuperconducting samples of $Na_{0.36(2)}CoO_{1.92(1)}$ (Sample B) and $Na_{0.36(2)}CoO_{1.92(1)} \cdot 0.5(1)\,H_2O$ (Sample D). This confirms previous suggestions that the proper oxidation state of cobalt alone does not facilitate superconductivity, but simultaneously required is proper spacing between the $CoO_2$ layers.[4,12] The most profound result however is that the value of $V$(Co) for Sample C, *i.e.* ~3.46, is markedly lower than ~3.64 expected on the basis of the Na content ($x$ = 0.36) only. The fact that $V$(Co) for this sample is even lower than 3.5 provides us with the basis to consider the superconducting $Na_{0.36(2)}CoO_{1.91(1)} \cdot 1.3(1)\,H_2O$ phase as a hole-doped $Co^{III}$ lattice. Note that in terms of the oxidation state of Co the $CoO_2$ layer in Sample C is not very far from that in the parent *p*-type thermoelectric compounds, $Na_xCoO_2$ ($0.57 \leq x \leq 0.75$)[3,11,15] having holes as charge carriers and $V$(Co) in the range of 3.25 to 3.43 (assuming oxygen stoichiometry). To confirm the type of charge carriers in the $Na_xCoO_{2-\delta} \cdot y\,H_2O$ superconductor Seebeck measurements are highly demanded.

The "lower-than-expected" $V$(Co) value for the superconducting $Na_xCoO_{2-\delta} \cdot y\,H_2O$ phase (Sample C) is compatible with the presence of oxygen vacancies with the concentration of $\delta \approx$ 0.09(1). For the $Na_xCoO_{2-\delta}$ system presence of oxygen vacancies is not a totally new fact, but already mentioned in the 1980s was that samples with reduced Na contents may be nonstoichiometric.[16,17] For a related system, $Li_xCoO_{2-\delta}$, quantitative data based on iodometric titration analysis were recently reported, showing that upon ambient-temperature extraction of Li using $NO_2PF_6$ in acetonitrile medium as an oxidant the phase started to deplete oxygen when the Li deficiency exceeded a level of (1-$x$) $\approx$ 0.35.[18,19] The maximum of $V$(Co) for $Li_xCoO_{2-\delta}$, *i.e.* ~3.45, appeared for $0.3 < x < 0.5$ with $\delta$ at $0.10 \sim 0.15$. For the fully-delithiated ($x$ = 0) metastable $CoO_{2-\delta}$ phase of the $CdI_2$ structure, $\delta$ and $V$(Co) were determined at ~0.33 and ~3.34, respectively.[18] Qualitatively, the presence of oxygen vacancies in the $CoO_{2-\delta}$ phase was also concluded from *in-situ* synchrotron x-ray diffraction data.[20] Recognizing the situation in $Li_xCoO_{2-\delta}$, the presently obtained results for the $Na_xCoO_{2-\delta} \cdot y\,H_2O$ superconductor appear highly plausible.

Assuming an analogy to the $CuO_2$-layer superconductors, it may sound somewhat amazing that the $CoO_2$ layer in $Na_xCoO_{2-\delta} \cdot y\,H_2O$ would be oxygen deficient and at the same time superconducting. Recently Rivadulla *et al.*[15] discussed the same subject hypothetically and proposed a model which assumes that - if existing - the $CoO_2$-layer oxygen vacancies would during the water intercalation be filled with the $O^{2-}$ heads of the intercalated water molecules. As a consequence, extra protons would be created in the structure. It may thus not be "$O^{2-}$ vacancies" but rather "excess $H^+$" to account for the "lower-than-expected" value of $V$(Co) for the superconducting phase, *i.e.* "$Na_xCoO_2 \cdot y\,H_2O_{1-\delta/y}$" or "$Na_xCoO_2 \cdot 2\delta H^+ \cdot (y-\delta)H_2O$". Keeping this in mind we look back at the TG curves shown in Figure 2. Among the $H_2O$ molecules those that have participated in filling the $CoO_2$-layer oxygen vacancies should be most strongly bound. Therefore they would be the last water molecules to be depleted from the structure upon heating. It is thus tempting to assign the very last step of weight loss seen in the TG curves at 200 - 230 $^o$C to these water molecules. This step corresponds to *ca.* 0.1 $H_2O$ molecules *per* formula unit, which well agrees with the estimated δ value of 0.09(1) for Sample C. However, we emphasize that the present study on the novel $Na_xCoO_{2-\delta} \cdot y\,H_2O$ superconductor which clearly concludes that the oxidation state of Co is lower than expected on the basis of stoichiometric oxygen content and the Na content being at $x \approx 0.36$, does not allow us to judge whether the obtained result is due to oxygen vacancies or excess protons. Here further careful studies by proper analytical tools are definitely required and warranted to clarify this point.

## Acknowledgments

Drs. E. Takayama-Muromachi and K. Takada are thanked for their initial advice in the synthesis of superconducting samples. The present work was supported by a grant from Hayashi Memorial Foundation for Female Natural Scientists and also by Grants-in-Aid for Scientific Research (Nos. 15206002 and 15206071) from the Japan Society for the Promotion of Science.

**Table 1.** Chemical characteristics of the present $Na_xCoO_{2-\delta} \cdot yH_2O$ samples, A-E, together with those of a "$Na_{0.75}CoO_{2-\delta}$ reference sample": Na content, $x$, from ICP analysis, water content, $y$, from TG analysis, $c$-axis lattice parameter from x-ray powder diffraction data, and the oxidation number of cobalt, $V(Co)$, and the oxygen nonstoichiometry, $\delta$, from cerimetric [iodometric] titration analysis. Note that only Sample C is superconducting ($T_c = 4.4$ K).

| Sample | $x$ | $y$ | $c$ [Å] | $V(Co)$ | $\delta$ |
|---|---|---|---|---|---|
| "$Na_{0.75}CoO_{2-\delta}$" | 0.77(2) | 0 | 10.92(1) | 3.26(1) [3.26(1)] | -0.02 [-0.02] |
| Sample A | 0.72(2) | 0 | 10.94(1) | 3.26(1) [3.25(1)] | 0.01 [0.01] |
| Sample B | 0.36(2) | 0 | 11.22(1) | 3.48(1) | 0.08 |
| Sample C | 0.36(2) | 1.3(1) | 19.64(1) | 3.48(1) [3.44(1)] | 0.08 [0.10] |
| Sample D | 0.36(2) | 0.5(1) | 13.83(1) | 3.48(1) | 0.08 |
| Sample E | 0.36* | 0 | 11.17(1) | 3.37(1) | 0.13 |

*used for the calculation of the values of $V(Co)$ and $\delta$ from the titration data

**Figure Captions**

**Figure 1.** Magnetic susceptibility ($\chi$) versus temperature as measured with a SQUID magnetometer in both field-cooled (FC) and zero-field-cooled (ZFC) modes for the Na-deintercalated, fully-hydrated $Na_xCoO_{2-\delta} \cdot y\, H_2O$ sample (Sample C) showing the superconductivity transition at 4.4 K.

**Figure 2.** Thermogravimetric curves for dehydration of the $Na_xCoO_{2-\delta} \cdot y\, H_2O$ superconductor (Sample C). The curves are for three parallel experiments carried out (each time) in $O_2$ for a ~20-mg specimen with a heating rate 0.25 $^o$C/min. Indicated are the temperatures used to synthesize Samples D, E and F from Sample C, and also the reflection point () defining the onset temperature for the very last water-depletion step from the $Na_{0.36}CoO_{1.91} \cdot y\, H_2O$ structure (see the text).

**Figure 3.** X-ray diffraction patterns for $Na_xCoO_2 \cdot y\, H_2O$ samples with different Na and $H_2O$ contents: Sample A ($Na_{0.72}CoO_{1.99}$ precursor), Sample B ($Na_{0.36}CoO_{1.91}$), Sample C ($Na_{0.36}CoO_{1.91} \cdot 1.3\, H_2O$ superconductor), Sample D ($Na_{0.36}CoO_{1.92} \cdot 0.5\, H_2O$), Sample E ($Na_{0.36}CoO_{1.87}$) and Sample F (a mixture of $Na_xCoO_{2-\delta}$ and $Co_3O_4$).

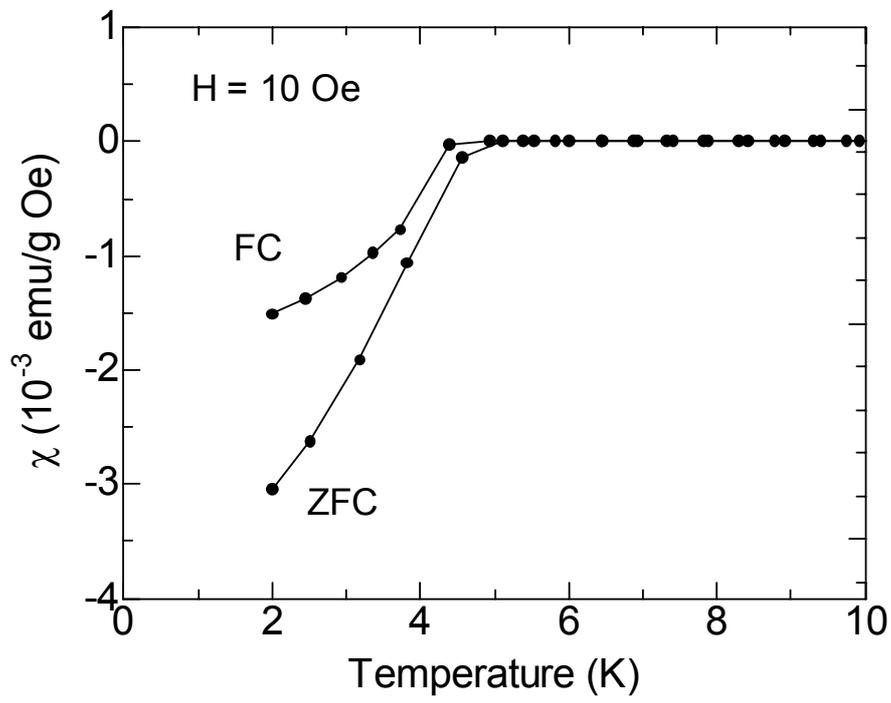

Karppinen *et al*.: Fig. 1.

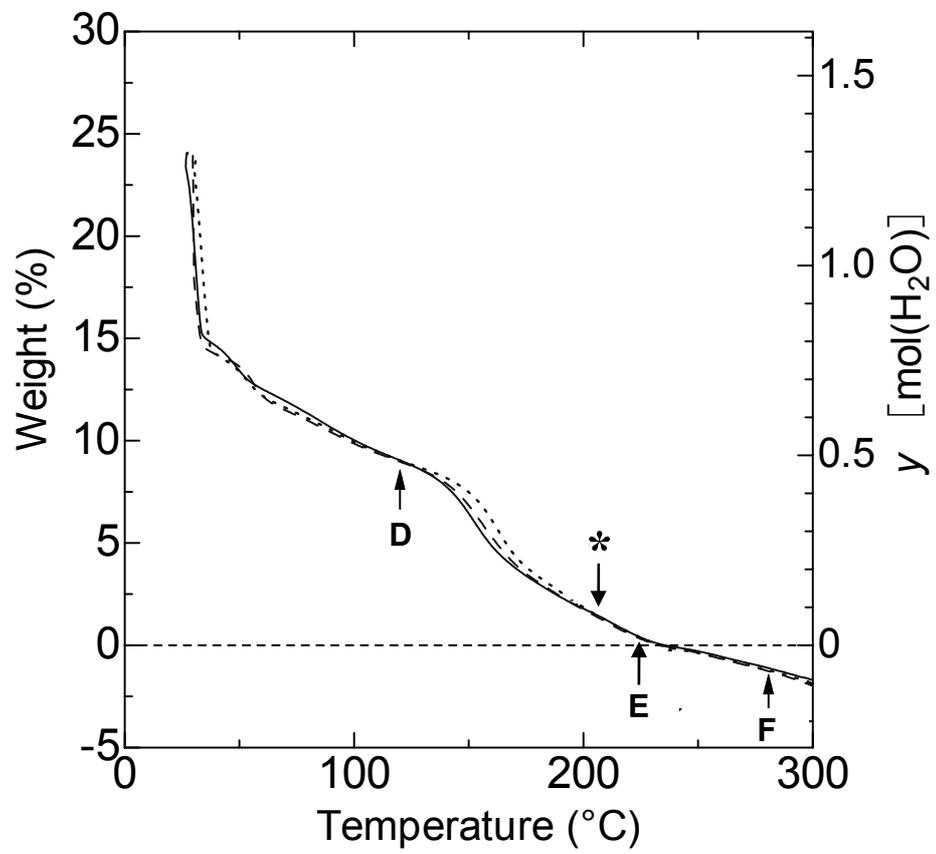

Karppinen *et al.*: Fig. 2.

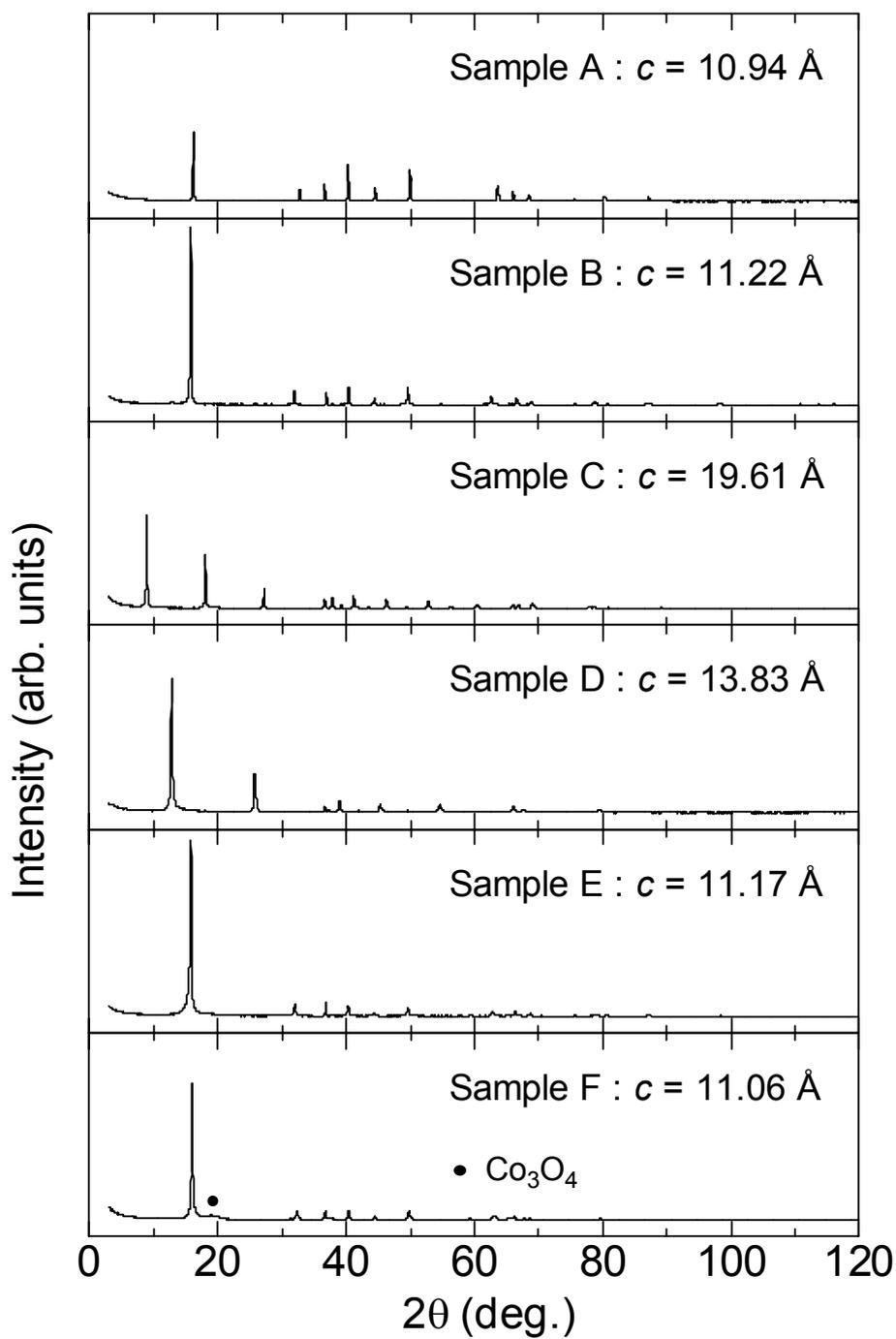

Karppinen *et al.*: Fig. 3.